\documentclass[%
superscriptaddress,
nofootinbib,
twocolumn,
amsmath,
amssymb, 
longbibliography
]{revtex4-1}

\usepackage{graphicx}
\usepackage{dcolumn}
\usepackage{bm}
\usepackage{braket}
\usepackage{color}

\usepackage[cal=esstix,frak=euler,scr=boondox,bb= pazo]{mathalfa}

\usepackage{float}
\begin{document}

\preprint{APS/123-QED}

\title{{Observing the escape of a driven quantum Josephson circuit into unconfined states}}

\author{Rapha\"el Lescanne}
\email{raphael.lescanne@lpa.ens.fr}
\affiliation{Laboratoire Pierre Aigrain, Ecole Normale Sup\'erieure, PSL Research University, CNRS, Universit\'e Pierre et Marie Curie, Sorbonne Universit\'es, Universit\'e Paris Diderot, Sorbonne Paris-Cit\'e, 24 rue Lhomond, 75231 Paris Cedex 05, France}
\affiliation{QUANTIC team, INRIA de Paris, 2 Rue Simone Iff, 75012 Paris, France}
\author{Lucas Verney}
\affiliation{QUANTIC team, INRIA de Paris, 2 Rue Simone Iff, 75012 Paris, France}
\affiliation{Laboratoire Pierre Aigrain, Ecole Normale Sup\'erieure, PSL Research University, CNRS, Universit\'e Pierre et Marie Curie, Sorbonne Universit\'es, Universit\'e Paris Diderot, Sorbonne Paris-Cit\'e, 24 rue Lhomond, 75231 Paris Cedex 05, France}
\author{Quentin Ficheux}
\affiliation{Laboratoire Pierre Aigrain, Ecole Normale Sup\'erieure, PSL Research University, CNRS, Universit\'e Pierre et Marie Curie, Sorbonne Universit\'es, Universit\'e Paris Diderot, Sorbonne Paris-Cit\'e, 24 rue Lhomond, 75231 Paris Cedex 05, France}
\affiliation{Laboratoire de Physique, Ecole Normale Sup\'erieure de Lyon, 46 all\'ee d'Italie, 69364 Lyon Cedex 7, France}
\author{Michel H. Devoret}
\affiliation{Department of Applied Physics, Yale University, 15 Prospect St., New Haven, CT 06520, USA}
\author{Benjamin Huard}
\affiliation{Laboratoire de Physique, Ecole Normale Sup\'erieure de Lyon, 46 all\'ee d'Italie, 69364 Lyon Cedex 7, France}
\author{Mazyar Mirrahimi}
\affiliation{QUANTIC team, INRIA de Paris, 2 Rue Simone Iff, 75012 Paris, France}
\affiliation{Yale Quantum Institute, Yale University, 17 Hillhouse Av., New Haven, CT 06520, USA}
\author{Zaki Leghtas}
\email{zaki.leghtas@mines-paristech.fr}
\affiliation{Centre Automatique et Syst\`emes, Mines-ParisTech, PSL Research University, 60, bd Saint-Michel, 75006 Paris, France}
\affiliation{Laboratoire Pierre Aigrain, Ecole Normale Sup\'erieure, PSL Research University, CNRS, Universit\'e Pierre et Marie Curie, Sorbonne Universit\'es, Universit\'e Paris Diderot, Sorbonne Paris-Cit\'e, 24 rue Lhomond, 75231 Paris Cedex 05, France}
\affiliation{QUANTIC team, INRIA de Paris, 2 Rue Simone Iff, 75012 Paris, France}

\date{\today}

\begin{abstract}


{Josephson circuits have been ideal systems to study complex non-linear dynamics which can lead to chaotic behavior and instabilities. More recently, Josephson circuits in the quantum regime, particularly in the presence of microwave drives,} {have demonstrated their ability} {to emulate a variety of Hamiltonians that are useful for the processing of quantum information. In this paper we show that these drives lead to an instability which results in the escape of the circuit mode into states that are not confined by the Josephson cosine potential. We observe this escape in a ubiquitous circuit: a transmon embedded in a 3D cavity. When the transmon occupies these free-particle-like states, the circuit behaves as though the junction ha{d} been removed, and all non-linearities are lost. This work deepens our understanding of strongly driven Josephson circuits, which is important for fundamental and application perspectives, such as the engineering of Hamiltonians by parametric pumping.}

\end{abstract}

\maketitle

\section{Introduction}
Superconducting circuits are one of the leading platforms to implement quantum technologies. They host highly coherent electromagnetic modes whose parameters are engineered to fulfill a particular function. Josephson junctions (JJ) mediate non linear couplings between the circuit modes which can be arranged in a variety of topologies. By tailoring the type and strength of coupling, one can perform a multitude of tasks, such as amplifying signals \cite{Castellanos-Beltran2008, Bergeal2010, Macklin2015}, generating non-classical light \cite{Pechal2014, Flurin2012}, stabilizing single quantum states \cite{Murch2012, Shankar2013} or manifolds \cite{Leghtas2015}, releasing and catching propagating modes \cite{Narla2016,Pfaff2017}, and simulating quantum systems \cite{Puri2017}. The full in-situ control of the couplings mediated by the Josephson non-linearity necessitates a key ingredient: a strong coherent microwave drive, referred to as a pump. Typically, the pump frequency is not resonant with any mode and instead satisfies a specific frequency matching condition in order to select the desired coupling (so-called parametric pumping). In the low pump power regime, {the system behaves in a } {stable} {manner, and} the coupling strength increases with the pump power.

{However, as the pump power is increased to seek stronger couplings, this non-linear out-of-equilibrium open quantum system can display complex dynamics leading to the degradation of coherence times~\cite{Leghtas2015,Gao-PRX_2018} and instabilities~\cite{Verney-al-inpreparation}. This behavior is {reminiscent} of the dynamics of RF current biased Josephson junctions, which have been thoroughly studied in the classical regime in the context of the Josephson voltage standard \cite{Kautz}. Understanding these dynamics in the quantum regime is essential, both from a fundamental and application perspective. A particularly urgent matter is to guide us towards circuit designs which prevent instabilities, making them suitable for parametric pumping.}


{This work focuses on the transmon~\cite{Koch2007}, a widely used superconducting circuit because of its long coherence times. {Usually,} its first two energy levels are used as a qubit to encode quantum information. Actually, the transmon has an infinite number of energy levels which fall in two categories. First, those with energies smaller than the Josephson energy and that are confined by the Josephson potential. Second, those which lie above the cosine potential which we refer to as unconfined states. Due to their high energies, unconfined states have always been considered irrelevant for circuit dynamics and disregarded \cite{Ginossar-al-PRA_2010, Bishop-al-PRL_2010, Boissonneault-al-PRL_2010, Mavrogordatos-PRL_2017, Sank-al-PRL_2016, Fink-al-PRX_2017}.}

{In this paper, we show that unconfined states play a central role in the dynamics of strongly driven Josephson circuits. This time-periodic system has periodic orbits known as Floquet states. In the dissipative steady state regime, the system converges to a statistical mixture of Floquet states. For weak pump powers, the steady state remains pure and occupies low energy confined states. Above a critical pump power, the steady state suddenly jumps to a statistical mixture of many Floquet states, with a significant population on unconfined states. The dramatic change in the steady state when the pump power is slightly increased above the threshold is a signature of structural instability \cite{Verney-al-inpreparation}}.
\section{Description of the experiment}
\begin{figure*}
\includegraphics[width=1.5\columnwidth]{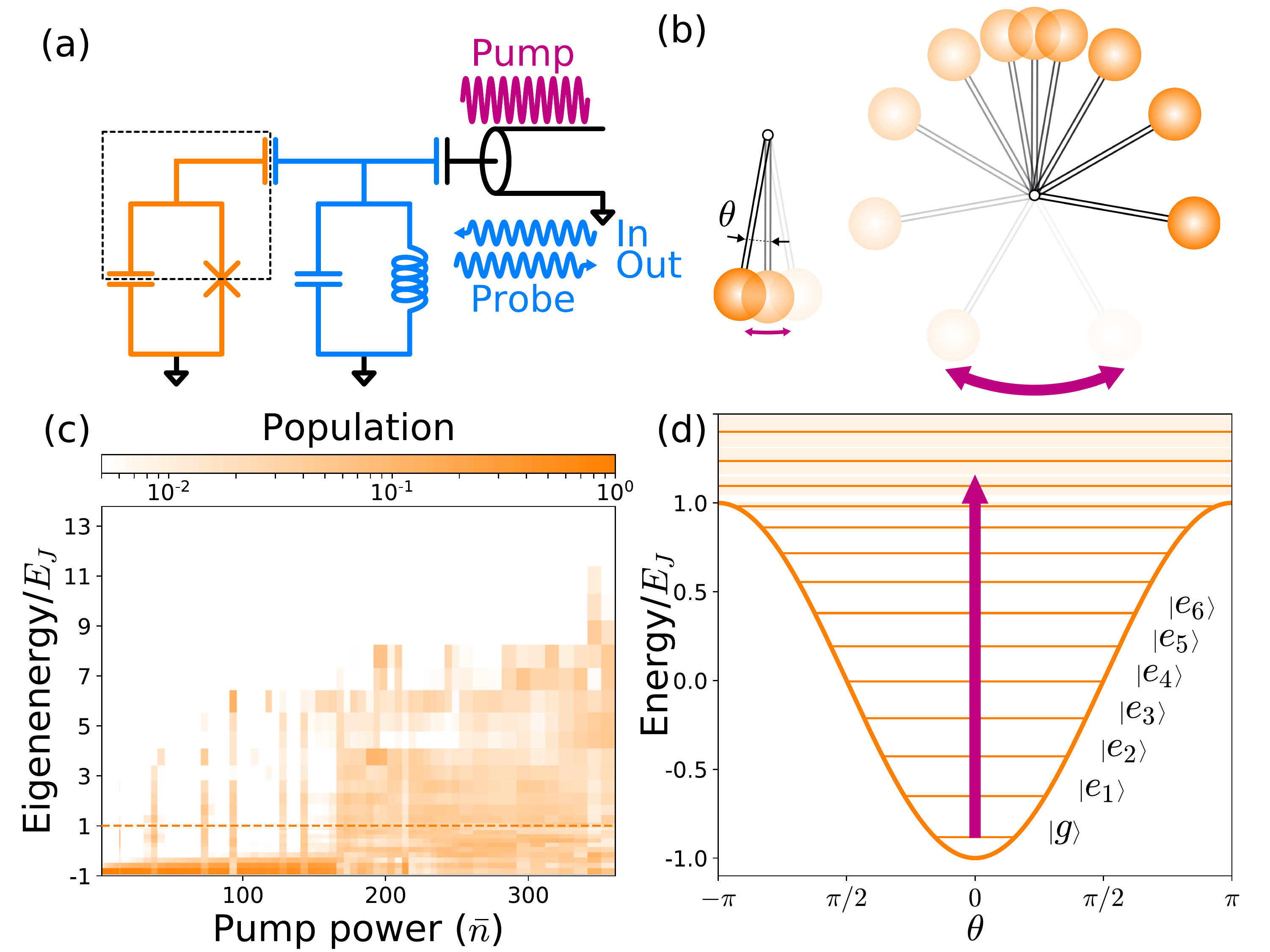}
 \caption{Principle of the experiment demonstrating the effect of a strong pump on a Josephson circuit. (\textbf{a}) A superconducting island (within the dashed lines) is coupled to ground through a Josephson junction (JJ), which is shunted by a capacitor, forming a transmon mode (orange). It is capacitively coupled to an LC resonator (blue) which represents the lowest mode of a 3D waveguide cavity. The system is pumped and probed through a transmission line (black). (\textbf{b}) Classical picture: The transmon is represented by a pendulum \cite{Koch2007}, where the deviation angle from equilibrium, denoted  $\theta$, is the phase difference across the JJ. The gravitational potential energy represents the Josephson energy. For small pump powers (left), the pendulum acquires a small kinetic energy and oscillates around its equilibrium. For large pump powers (right), it acquires sufficient energy to escape from its trapping potential well, and rotates indefinitely. (\textbf{c}) Numerical simulation of the population (color) of the energy levels of the transmon (y-axis) as a function of the number of photons $\bar{n}$ (Appendix~\ref{appendix:phnumcal}) in the resonator due to the pump (x-axis). For small $\bar{n}$, the transmon is close to its ground state, except for small population bursts (see text). Above a critical $\bar{n}$ (here about 170 photons), the transmon jumps into highly excited states above its cosine potential (orange dashed line). (\textbf{d}) Quantum picture: cosine potential and energy levels (orange lines) for charge offset $N_g=0$. {Light orange} areas represent the band of levels appearing when varying $N_g$ \cite{cottetthesis}. Large pump powers (big arrow) induce transitions from the ground state $\ket{g}$ with a phase localized around $\theta = 0$, into highly excited states with a delocalized phase.}
\label{fig1}
\end{figure*}

Our device consists of a single transmon in a 3D copper cavity \cite{Paik2011} coupled to a transmission line. The system is well modeled by the circuit depicted in Fig.~\ref{fig1}a, whose Hamiltonian is given by \cite{Koch2007}
\begin{eqnarray}
\label{hamiltonian}
\bm{H}(t) & = & 4E_C \bm{N}^2-E_J\cos\left(\bm{\theta}\right)
+\hbar\omega_a\bm{a}^\dag \bm{a}
+ \hbar g\bm{N}\left(\bm{a}+\bm{a}^\dag\right)\notag\\ &+& \hbar \mathcal{A}_p(t)\left(\bm{a}+\bm{a}^\dag\right),
\end{eqnarray}
where $\bm{N}$ is the Cooper pair number operator and $\bm{\theta}$ is the phase operator \cite{Girvin2011}. The operator $\cos\left(\bm{\theta}\right)$ is the transfer operator for Cooper pairs across the junction while $\bm{a}^\dag$ and $\bm{a}$ are the creation and annihilation operators for the resonator mode. 
The Josephson and charging energies are denoted $E_J$ and $E_C$, respectively. 
The angular frequency of the resonator in the absence of the JJ ($E_J$ taken to be 0) is denoted $\omega_a$ and is known as the bare resonator frequency and $g$ is the coupling rate between the transmon and the resonator. Note that this definition of $g$ differs from the one usually used in the Jaynes-Cummings Hamiltonian \cite{Wallraff2004}. The pump couples capacitively to the circuit, and is accounted for by the second line in \eqref{hamiltonian}, where $\mathcal{A}_p(t)=A_p\cos(\omega_p t)$, $A_p$ and $\omega_p$ being the pump amplitude and angular frequency. 

Our physical picture of the dynamics of a pumped transmon in a cavity is the following. When the pump power is sufficiently large, the transmon is excited above its bounded potential well (Fig.~\ref{fig1}c,d). This is {reminiscent} to an electron escaping from the bound states of the atomic Coulomb potential and leaving the atom ionized. The transmon can occupy highly excited states which resemble charge states, or equivalently, plane waves in the phase representation. A classical analogy would be a strongly driven pendulum which rotates indefinitely, making turns around its anchoring point (Fig.~\ref{fig1}b). When occupying such a state, the transmon behaves like a free particle which is almost not affected by the cosine potential. Thus the Josephson energy can be set to zero in Eq.~\eqref{hamiltonian}. The Hamiltonian is now that of a superconducting island capacitively coupled to a resonator, with no Josephson junction. The island-resonator coupling term commutes with the Hamiltonian of the island, which consists of the charging energy only. The coupling is now longitudinal and therefore the cavity frequency is fixed to the bare frequency $\omega_a$, independently of the island state.

The dynamics governed by Hamiltonian \eqref{hamiltonian} is difficult to numerically simulate in the regime of large pump amplitudes for two reasons. First, the simulated Hilbert space needs to be large. Indeed, the number of transmon states needs to be much larger than the number of states in the well. For our experimental parameters we find that about 45 states are {necessary}. Moreover, the resonator is also driven to large photon number states. Second, the pump amplitude is so large that we reach regimes where $A_p \gg \omega_a$. It is therefore excluded to use the rotating wave approximation in its usual form.
These two difficulties are surmounted by the theory introduced in \cite{Verney-al-inpreparation}. Using an adequate change of frame in which the resonator is close to the vacuum, we can decrease the number of computed states to 10 for the resonator, hence reducing the dimensionality of the Hilbert space (Appendix~\ref{appendix:numsim}). Since $\bm{H}(t)$ is $2\pi\omega_p^{-1}$-periodic, Floquet-Markov theory \cite{Grifoni-Hanggi-PhysRep_1998, Qutip} is well suited \cite{Pietikainen-PRB_2017,Pietikainen-JLTP_2018} to find the system steady state, as shown in Fig.~\ref{fig1}c. This theory assumes weak coupling to the bath, but can treat pumps of arbitrary amplitude and frequency. For small pump powers, the transmon remains close to its ground state, and after the power is increased beyond a critical value, it is driven into highly excited states, above its cosine potential (orange dashed line in Fig.~\ref{fig1}c). 

The parameters of Hamiltonian \eqref{hamiltonian} are deduced from measurements involving a low number of excitations in the system. In this regime, it is well approximated by the Hamiltonian of two non-linearly coupled effective modes, one transmon qubit-like and one resonator-like, respectively denoted by subscripts $q$ and $r$ \cite{Nigg2012}:
\begin{eqnarray}
\label{hamiltonian_low}
\bm{H}_\text{low}(t)\slash \hbar &=& \bar\omega_q \bm{a}_q^\dag\bm{a}_q - \frac{\alpha_q}{2} \left(\bm{a}_q^\dag\right)^2{\bm{a}_q}^2
- \chi_{qr} \bm{a}_q^\dag\bm{a}_q \bm{a}_r^\dag\bm{a}_r\notag\\ 
&+&\bar\omega_{r} \bm{a}_r^\dag\bm{a}_r - \frac{\alpha_r}{2} \left(\bm{a}_r^\dag\right)^2{\bm{a}_r}^2\notag\\ 
&+& \mathcal{A}_{p,r}(t)\left(\bm{a}_r+\bm{a}_r^\dag\right),
\end{eqnarray}
where $ \bar\omega_{q,r}$ are the modes angular frequencies, $\alpha_{q,r}$ their Kerr non-linearities, and $\chi_{qr}$ their dispersive coupling. In this basis, the pump is represented by the term in $\mathcal{A}_{p,r}(t)=A_{p,r}\cos(\omega_p t)$, with a renormalized amplitude $A_{p,r}$, where we neglect the linear coupling of the pump to mode $q$.

We measure $\bar\omega_q\slash 2\pi = 5.353$ GHz, $\bar\omega_r\slash 2\pi = 7.761$ GHz, $\alpha_q\slash 2\pi = 173$ MHz and $\chi_{qr}\slash 2\pi = 5$ MHz. In order to compute $E_J,~E_C,~g$ and $\omega_a$, we write Hamiltonian \eqref{hamiltonian} in matrix form in the charge basis for the transmon and the Fock basis for the resonator. We calculate the energy differences of the lowest energy levels, and fit the measured values with: ${E_C}/h = 166 ~\mathrm{MHz}$, ${E_J}/h = 23.3 ~\mathrm{GHz}$, ${g}/{2\pi} = 179 ~\mathrm{MHz}$ and $\omega_a/ 2\pi = 7.739~\mathrm{GHz}$. We also infer $\alpha_r/ 2\pi = 43~$kHz. 


\section{Observing the transmon escape into unconfined states}
\begin{figure*}
  \includegraphics[width=1.5\columnwidth]{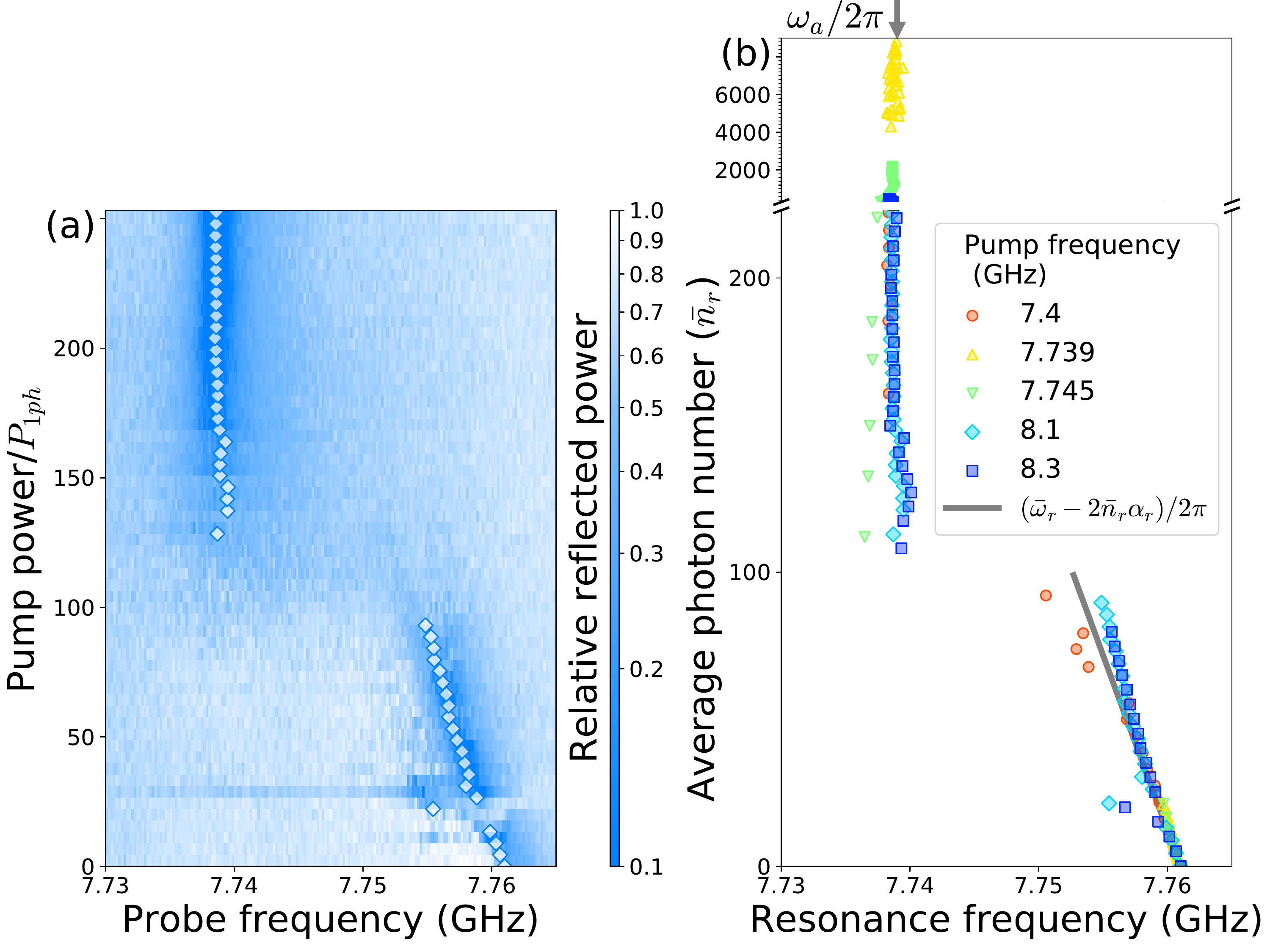}
 \caption{Effect of the pump on the cavity resonance frequency. (\textbf{a}) Relative reflected power of a weak probe as a function of probe frequency (x-axis), and pump power (y-axis) in units of $P_{1ph}$, the power needed to populate the cavity with one photon in average. The pump frequency is fixed at $8.1~$GHz, about $300~$MHz above the cavity frequency. The reduced reflected power is due to internal losses when the probe is resonant with the cavity. We indicate the fitted cavity resonance $\omega_r/2\pi$ with white diamonds. As the pump power increases, two regimes are distinguishable. For small powers, $\omega_r$ shifts linearly with the pump power. Above a critical power, the cavity resonance jumps to a new frequency $\omega_a/2\pi = 7.739$~GHz which is independent of pump power. (\textbf{b}) Fitted cavity resonance as a function of pump power for various pump frequencies. Pump powers are converted into a mean number of photons $\bar{n}_r$ (Appendix~\ref{appendix:phnumcal}). The y-axis of (a) and (b) slightly differ because $\bar{n}_r$ takes into account the resonator frequency shift. The general behavior is the same for all pump frequencies. The low power linear dependence is well reproduced by the AC Stark shift \cite{Leghtas2015} for an independently measured Kerr $\alpha_r$ (solid gray line).}
\label{fig2}
\end{figure*}

We experimentally observe the transmon jump into highly excited states by performing the spectroscopy of the resonator while the pump is applied (Fig.~\ref{fig2}). For small pump powers, the system is well described by the low energy Hamiltonian \eqref{hamiltonian_low}. The pump populates the resonator with photons at the pump frequency with an average number $\bar{n}_r=\left|\frac{A_{p,r}/2}{i(\omega_r-\omega_p)+\kappa_r\slash 2}\right|^2$, and shifts the resonance to $\omega_r(\bar{n}_r) = \bar\omega_r - 2\alpha_r \bar{n}_r$ \cite[Supplement section 2.1]{Leghtas2015}. We calibrate the photon number $\bar{n}_r$ (Appendix~\ref{appendix:phnumcal}) and, using the anharmonicity $\alpha_r$ deduced from the model \eqref{hamiltonian_low}, we plot $\omega_r(\bar{n}_r)$ (solid line in Fig.~\ref{fig2}b). The data points match this prediction in the regime of small photon numbers, over a wide range of pump frequencies. This linear dependence is disrupted by abrupt frequency jumps, as seen for a pump power of about $30P_{1ph}$ in Fig.~\ref{fig2}a. This behavior suggests that the system frequencies, at this specific power, have shifted into resonance with a process which excites the transmon. This resembles the population bursts observed for example at 90 photons in the simulation of Fig.~\ref{fig1}c, before the population jumps at around 170 photons.

Above a critical $\bar{n}_r$ (of about 100 photons for the most detuned pump frequencies), the resonance jumps to a new frequency close to $\omega_a\slash 2\pi$ (arrow in Fig.~\ref{fig2}b), which is independent of the pump frequency and power. This is the resonator frequency expected when the transmon is highly excited so that the JJ behaves as an ``open". Such a jump of the resonance has been {used for qubit readout \cite{Reed-PRL_2010} and} modeled in previous works with a strong {drive which plays the role of the pump and the probe}, applied close to resonance with the resonator. Those models involve two-level \cite{Bishop-al-PRL_2010} and multi-level \cite{Boissonneault-al-PRL_2010} approximations for the transmon. Such approximations are incompatible with our following measurements. In recent work \cite{Pietikainen-PRB_2017,Pietikainen-JLTP_2018}, numerical simulations suggest that the resonator frequency jump is accompanied by an increase in the transmon population. In the next paragraphs, we experimentally measure a sudden population increase which corresponds to the jump of the transmon state out of its potential well into unconfined states. In order to model this effect, it is crucial to consider all levels confined in the cosine potential and a number of levels above which depends on the pump power.

\begin{figure*}
\includegraphics[width=1.5\columnwidth]{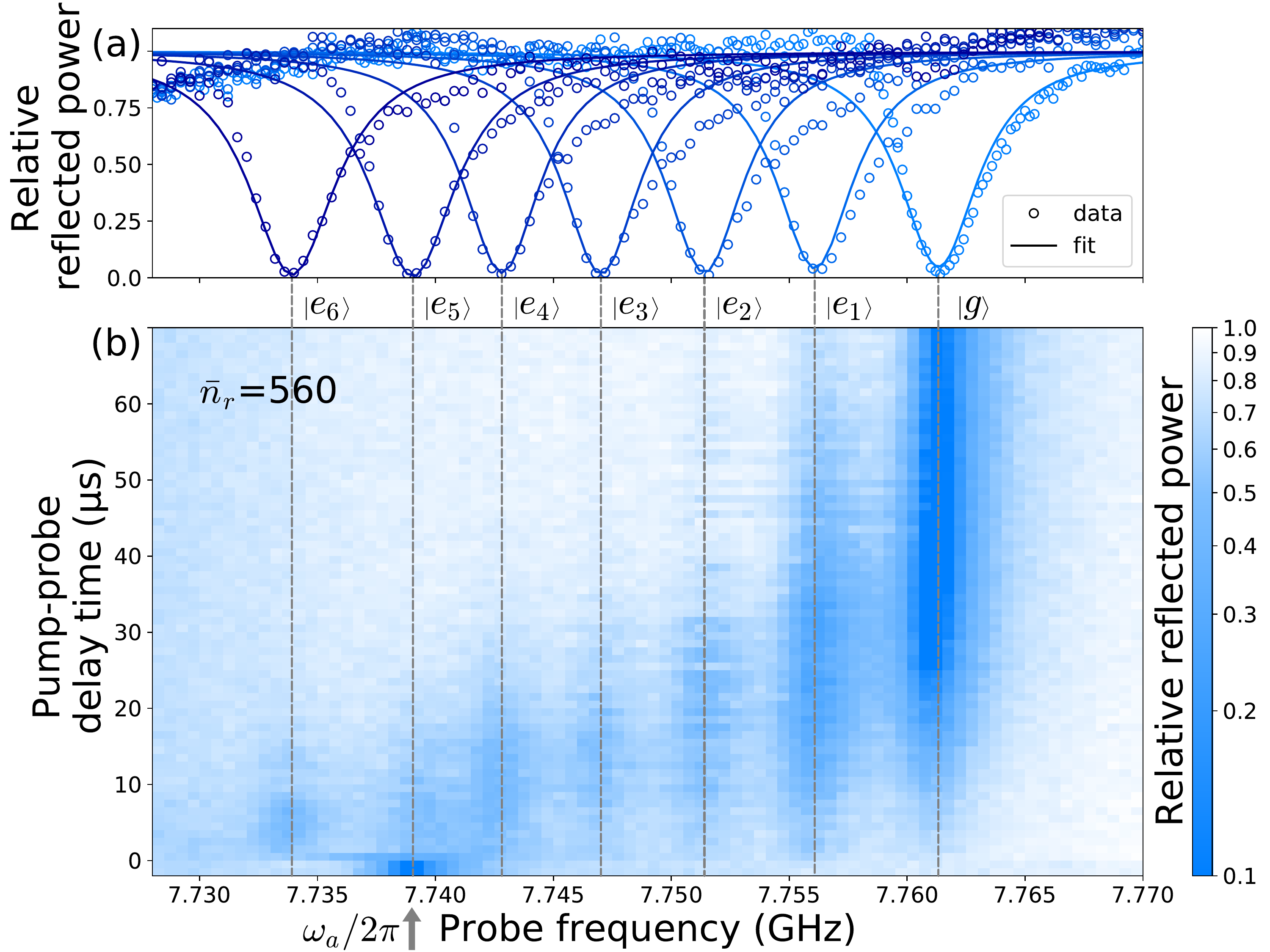}
\caption{Probing the transmon decay out of highly excited states populated by a strong off-resonant pump. (\textbf{a}) Cavity spectroscopy after the transmon is prepared in various eigenstates.  Each transmon state $\ket{e_k}$ ($k=$1 to 6) is prepared using a $k$-photon $\pi$-pulse (Fig.~\ref{figs2}) and higher states could not be prepared due to charge noise (Appendix \ref{appendix:chargenoise}). (\textbf{b}) Time resolved measurement: first, a pump is applied for $50~\mu$s at $8.1~$GHz populating the cavity with a sufficiently large photon number to induce a jump in $\omega_r$, here $\bar{n}_r=560$. After a time-delay $t$, a weak probe is applied for $2~\mu$s. We plot the relative reflected power of the probe as a function of probe frequency (x-axis), and time-delay $t$ (y-axis). For $t<0$, as in Fig.~\ref{fig2}, the cavity is probed while the pump is on,  and the resonance frequency is $\omega_a/2\pi$, confirming that the system has jumped. At $t>0$, the photons in the cavity rapidly decay at a rate $\kappa_r=1/(55~\text{ns})$, and the cavity can now be used, as in Fig.~\ref{fig3}a, to read the transmon state.  The transmon is found in a highly excited state, and after $t=60~\mu$s, it has fully decayed into its ground state which is compatible with $T_1=14~\mu$s.}
\label{fig3}
\end{figure*}

For states well confined inside the cosine potential, the transmon state is encoded in the resonance frequency of the resonator. As seen from Hamiltonian \eqref{hamiltonian_low}, each additional excitation in the transmon pulls the resonator frequency to $\omega_r(n_q) = \bar\omega_r - n_q\chi_{qr}$ (Fig.~\ref{fig3}a). However, for states above the cosine potential, the resonator adopts a frequency $\omega_a$ which is now independent of the particular transmon state.

The highly excited nature of the transmon can be confirmed by turning off the pump, and waiting for the transmon to decay to lower lying states which can be resolved by probing the resonator frequency. In this experiment, the highest of these states was the sixth excited state $\ket{e_6}$. As expected, after the pump is turned off ($t=0$ in Fig.~\ref{fig3}b), we observe a population decay of the transmon compatible with a highly excited state at $t\le 0$.



\section{Effect of pump-induced quasiparticles}
\label{sec:quasiparticles}
\begin{figure*}
\includegraphics[width=1.5\columnwidth]{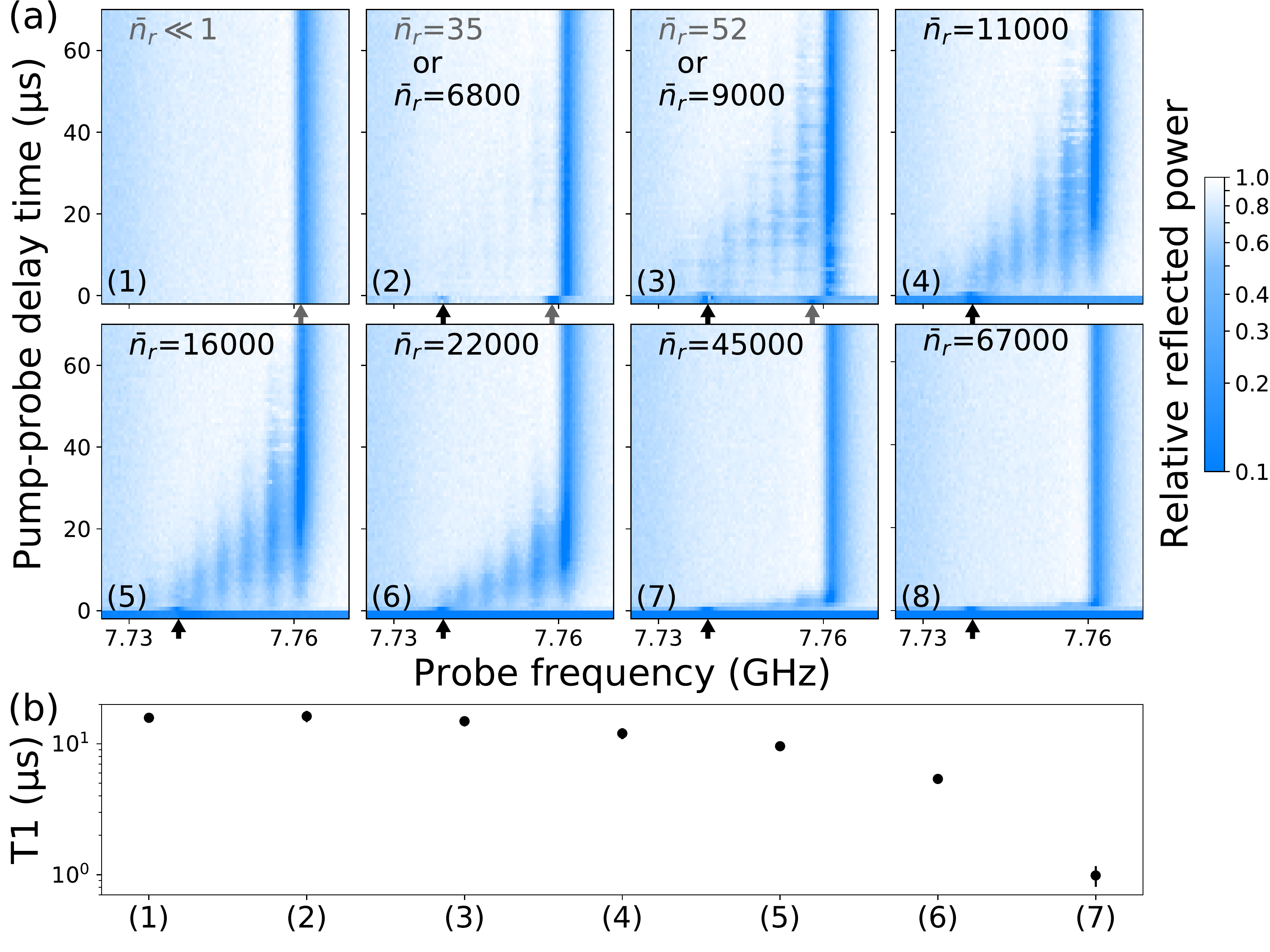}
\caption{Transmon relaxation following a pump pulse of increasing power. (\textbf{a}) Same experiment as in Fig.~\ref{fig3}b with increasing pump power (a1 to a8). The pump frequency is fixed at $f_p = \omega_a/2\pi = 7.739$ GHz in order to populate the cavity with a large number of photons. Each pump power is converted into the mean number of photons $\bar{n}_r$ in the cavity. The conversion factor depends on the observed cavity frequency at $t\leq 0$, indicated by the arrows (Appendix \ref{appendix:phnumcal}). (\textbf{a1}) The pump power is too low to induce a shift or a jump in the cavity frequency. (\textbf{a2}-\textbf{a3}) At intermediate pump powers, the system is in a statistical mixture of two configurations, as indicated by the two arrows at $t<0$. First, a weakly shifted resonance due to the AC Stark shift with a small cavity population. Second, a resonance at $\omega_a$ coinciding with the pump frequency and hence a large cavity population. (\textbf{a4}-\textbf{a5}) At higher pump powers, the cavity frequency always jumps to $\omega_a$ and the transmon decays as in Fig.~\ref{fig3}b. (\textbf{a6}-\textbf{a8}) At even higher pump powers, the transmon decay rate increases significantly. (\textbf{b}) Transmon $T_1$ (y-axis) measured 50$~\mu$s after a pump is applied. The pump duration and powers (x-axis) correspond to the ones used for the experiments shown in (a). We observe a decrease in $T_1$ as the pump power increases, which is consistent with an increasing quasi-particle density induced by the pump \cite{Wang-al-NatureComm_2014, Vool-al-PRL_2014}, and explains the significantly higher decay rates in (a6-a8). For pump power (3), $T_1$ is marginally reduced, while the system dynamics is highly affected (a3). The $T_1$ corresponding to (a8) was too small to be accurately measured.
}
\label{fig4}
\end{figure*}

Strongly pumping a superconducting circuit generates non-equilibrium quasiparticles \cite{Wang-al-NatureComm_2014, Vool-al-PRL_2014}, which poses the question: what is the influence of quasiparticles on this dynamics?
We detect whether quasiparticles have been generated by the pump pulse by measuring the transmon $T_1$ after a time delay large enough to ensure that all modes have decayed back to their ground states. We find that for the pump frequency and power used for the experiment of Fig.~\ref{fig3}b, the $T_1$ is unaffected (Fig.~\ref{figs4}). This indicates that the pump can expel the transmon out of its potential well, without generating a measurable amount of quasiparticles, and therefore the simulation of Fig.~\ref{fig1}c, which does not include quasiparticles, is relevant.

In order to observe a measurable decrease in $T_1$, we need to insert a much larger number of photons in the resonator. We enter this regime by applying a pump at frequency $\omega_a$, so that the pump is resonant with the resonator at large powers. As shown in Fig.~\ref{fig4}b, the $T_1$ drops at high pump powers indicating that we have successfully generated quasiparticles. Furthermore, we find that the transmon $T_1$ recovers its nominal value after a few milliseconds, a typical timescale for quasiparticle recombination \cite{Wang-al-NatureComm_2014, Vool-al-PRL_2014}. In Fig.~\ref{fig4}a, we probe the transmon dynamics as in Fig.~\ref{fig3}b, in the presence of pump-induced quasiparticles. The increased quasiparticle density does not qualitatively modify the transmon dynamics, but only increases the decay rate back to its ground state.


\section{Conclusion}
In conclusion, we have measured the dynamics of a transmon in a cavity in the presence of a strong off-resonant pump. Our measurements are in qualitative agreement with our theory describing the transmon escaping out of its potential well. The transmon then behaves like a free particle which no longer induces any non-linearity on the cavity. This escape, which is due to the boundedness of the cosine potential, could be prevented by adding an unbounded parabolic potential, provided by an inductive shunt. One could then further increase the non-linear couplings by increasing the pump power, without driving the transmon into free particle states and losing the resource of non-linearity. Our research points towards inductively shunted JJs \cite{Koch-PRL_2009} as better suited than standard transmons for pump-induced non-linear couplings \cite{Verney-al-inpreparation}.

\section{acknowledgments} 
The authors thank Pierre Rouchon, Takis Kontos, Benoit Dou\c{c}ot and \c{C}a\u{g}lar Girit for fruitful discussions and Matthieu Dartiailh for developing the instrument control software \cite{exopy}. This work was supported by the ANR grant ENDURANCE, and the EMERGENCES grant ENDURANCE of Ville de Paris. The devices were fabricated within the consortium Salle Blanche Paris Centre. MHD acknowledges ARO support.

\section{Appendix}

\subsection{Sample fabrication}
\label{appendix:fab}
The sample measured in this work is the same one as in \cite{Ficheux2018}. The transmon is made of a single aluminum Josephson junction and is embedded in a copper cavity of $26.5 \times 26.5 \times 9.5 \mathrm{~mm^3}$ thermalized on the base plate of a dilution fridge at about $10 \mathrm{~mK}$. More details can be found in \cite[Sections 1.A and 1.B]{Ficheux2018}.

\subsection{Photon number calibration}
\label{appendix:phnumcal}
\begin{figure*}
\includegraphics[width=1.8\columnwidth]{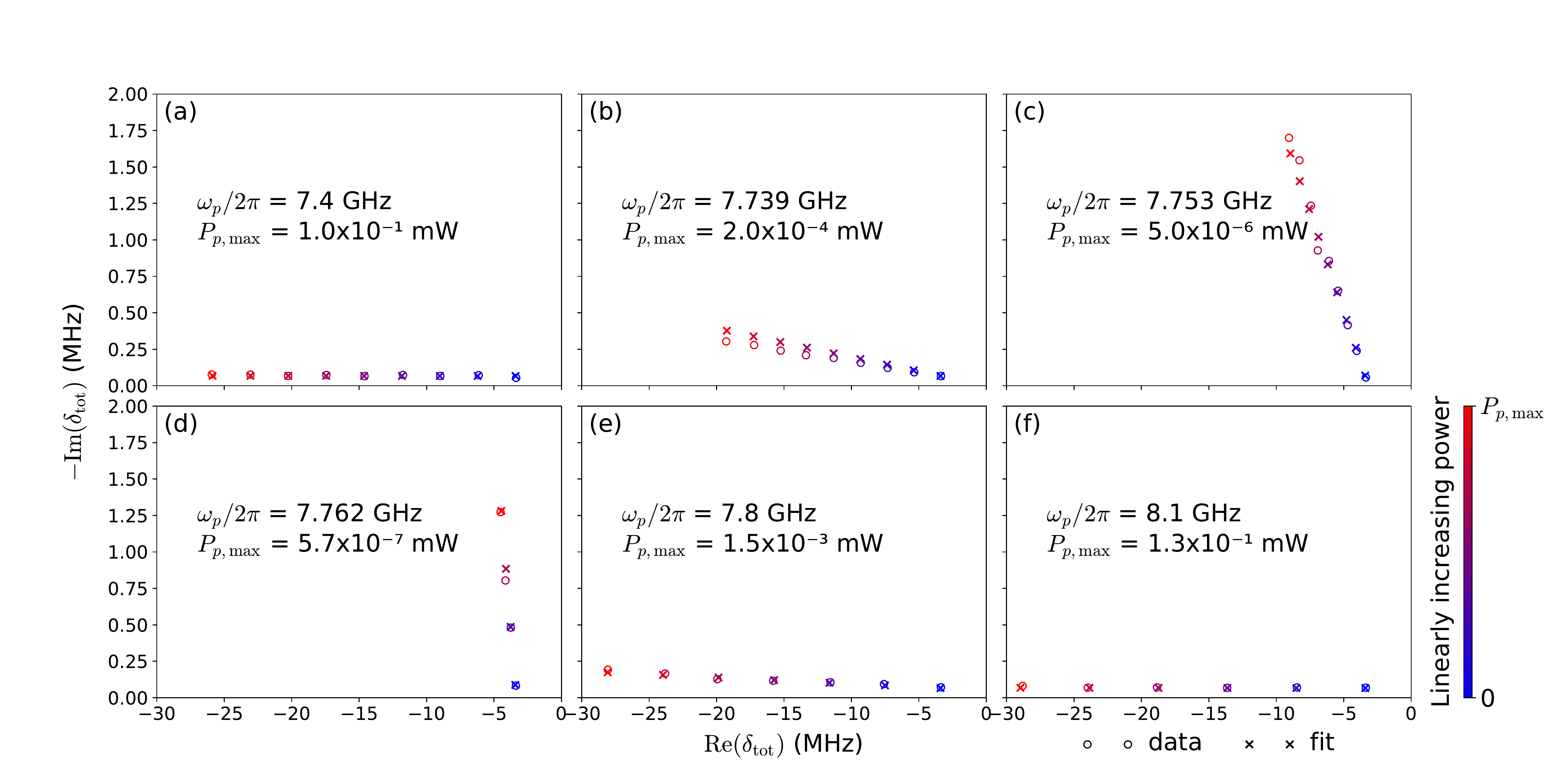}
  \caption{Calibration to convert the pump power (mW) into an average number of photons $\bar{n}_r$, using the AC Stark shift and measurement-induced dephasing \cite{Gambetta2006}. By performing a Ramsey experiment in the presence of a pump at frequency $\omega_p$ and power $P_p$, we measure the detuning between the drive and the qubit $\delta_q(\omega_p, P_p)$, and the qubit dephasing rate $\gamma_\phi(\omega_p, P_p)$. We introduce $\delta_{tot} = \delta_q - i\gamma_\phi$. For each plot (a) to (f), the pump frequency is fixed, and $\gamma_{tot}$ is plotted in the complex plane (hollow circles), for increasing $P_p$ (blue is $0~$mW, red is $P_{p,\text{max}}$). For each pump frequency, the data points are fitted with Eq.~\eqref{deltatot}, where $\delta_m$ is given in Eq.~\eqref{deltam}, and the constant $C$ is the only fitting parameter. All other parameters: $\Delta, \kappa_r, \chi_{qr}, P_p, \delta_0$ are measured independently. (a),(f) Large detuning between the pump and the cavity. The pump mostly induces a detuning on the qubit, without inducing any dephasing. (d) The pump is resonant with the cavity, and mostly induces dephasing on the qubit. (b), (c), (e) Intermediate regimes where the pump both dephases and detunes the qubit. }
\label{figs1}
\end{figure*}

In this section, we explain how we convert a pump power in mW, to an average number of photons at the pump frequency inside the cavity. We exploit the AC Stark shift and measurement induced-dephasing \cite{Gambetta2006}, where a drive on the resonator inserts photons which detune and dephase the qubit. We use a 2-level approximation of the low energy Hamiltonian from Eq.~\eqref{hamiltonian_low}: 
\begin{eqnarray}
\bm{H}^{(2)}_\text{low} \slash \hbar &=&  \omega_q\frac{\bm{\sigma}_z}{2} + \omega_r \bm{a}_r^\dag \bm{a}_r - \chi_{qr}\ket{e}\bra{e} \bm{a}_r^\dag \bm{a}_r  \notag\\
&+&A_{p,r}\cos(\omega_p t)\left(\bm{a}_r+\bm{a}_r^\dag\right)
\end{eqnarray}
where $\sigma_z=\ket{e}\bra{e}-\ket{g}\bra{g}$, and $\ket{g}, \ket{e}$ are the transmon's ground and first excited states. Moving into a frame rotating at the pump frequency $\omega_p$ for the resonator, and the qubit frequency $\omega_q$ for the qubit, and performing the rotating wave approximation (RWA), we get
\begin{eqnarray*}
\bar{\bm{H}}^\text{(2)}_\text{low} \slash \hbar &=& \Delta \bm{a}_r^\dag \bm{a}_r - \chi_{qr}\ket{e}\bra{e} \bm{a}_r^\dag \bm{a}_r \\
&+& A_{p,r} \bm{a}_r \slash 2 + A_{p,r}\bm{a}_r^\dag \slash 2\;,
\end{eqnarray*}
where $\Delta = \omega_r - \omega_p$.

Depending on the qubit state $\ket{g}$ [resp: $\ket{e}$], the resonator reaches a steady state which is a coherent state of amplitude $\alpha_g$ [resp: $\alpha_e$], where:
\begin{eqnarray*}
\alpha_g &=& \frac{-iA_{p,r}\slash 2}{i\Delta + \kappa_r\slash 2}\\
\alpha_e &=& \frac{-iA_{p,r}\slash 2}{i(\Delta-\chi_{qr}) + \kappa_r\slash 2}\;.
\end{eqnarray*}
The measurement induced complex detuning is \cite{Gambetta2006}
\begin{eqnarray*}
\delta_m &=& -\chi_{qr}\alpha_e\alpha_g^*\\
&=& -\chi_{qr}\frac{A_{p,r}^2\slash 4}{\left(i(\Delta-\chi_{qr}) + \kappa_r\slash 2\right)\left(-i\Delta + \kappa_r\slash 2\right)}\;,
\end{eqnarray*}
Re($\delta_m$) being the AC Stark frequency shift, and -Im($\delta_m$) being the measurement induced dephasing rate. The quantity $A_{p,r}^2$ is proportional to the output power of our instrument $P_p$, and hence we write it in the following form: $A_{p,r}^2 = \kappa_r^2 C P_p$, where $C$ is the proportionality constant we aim to calibrate and is expressed in a number of photons per mW. Since our microwave input lines do not have a perfectly flat transmission over a GHz range, $C$ can be dependent on $\omega_p$. We get:

\begin{equation}
\label{deltam}
\delta_m(\omega_p, P_p)
= -C\frac{\kappa_r^2\chi_{qr}P_p\slash 4}{\left(i(\Delta-\chi_{qr}) + \kappa_r\slash 2\right)\left(-i\Delta + \kappa_r\slash 2\right)}\;.
\end{equation}

For each pump frequency $\omega_p$, we vary the pump power $P_p$ between 0 and $P_{p,max}$, and perform a $T_2$ Ramsey experiment. The Ramsey fringes oscillation frequency and exponential decay provide a complex frequency shift $\delta_\text{tot}$, such that Re$(\delta_\text{tot})$ is the oscillation frequency, and -Im$(\delta_\text{tot})$ is the dephasing rate. We have
\begin{equation}
\label{deltatot}
\delta_\text{tot}(\omega_p, P_p) = \delta_{0}+\delta_m(\omega_p, P_p)\;,
\end{equation}
where $\delta_0$ is due to the chosen drive-qubit detuning and to the qubit dephasing rate in the absence of the pump. Thus, we measure $\delta_\text{tot}$, and the only unknown is the constant $C$, which we fit at each pump frequency. In Fig.~\ref{figs1}, we plot for each pump frequency, $\delta_\text{tot}$ (open circles), and the fitted values (crosses), where $C$ is the only fitting parameter. Once we know $C$, the average photon number is given by

\begin{eqnarray}
\bar{n}_r(\omega_p, P_p) &=& \left|\frac{-iA_{p,r}\slash 2}{i(\omega_r-\omega_p) + \kappa_r\slash 2}\right|^2\\
\label{nbar_r}
&=&CP_p\frac{\kappa_r^2\slash 4}{(\omega_r-\omega_p)^2 + \kappa_r^2\slash 4}\;,
\end{eqnarray}
where $\omega_r$ is the measured resonance frequency which itself depends on $\omega_p$ and $P_p$. We use Eq.~\eqref{nbar_r} to calculate all the given values of $\bar{n}_r$, and for the y-axis of Fig.~\ref{fig2}b.

\subsection{multi-photon qubit drive}
\label{appendix:multiphotondrive}
\begin{figure*}
\includegraphics[width=1.8\columnwidth]{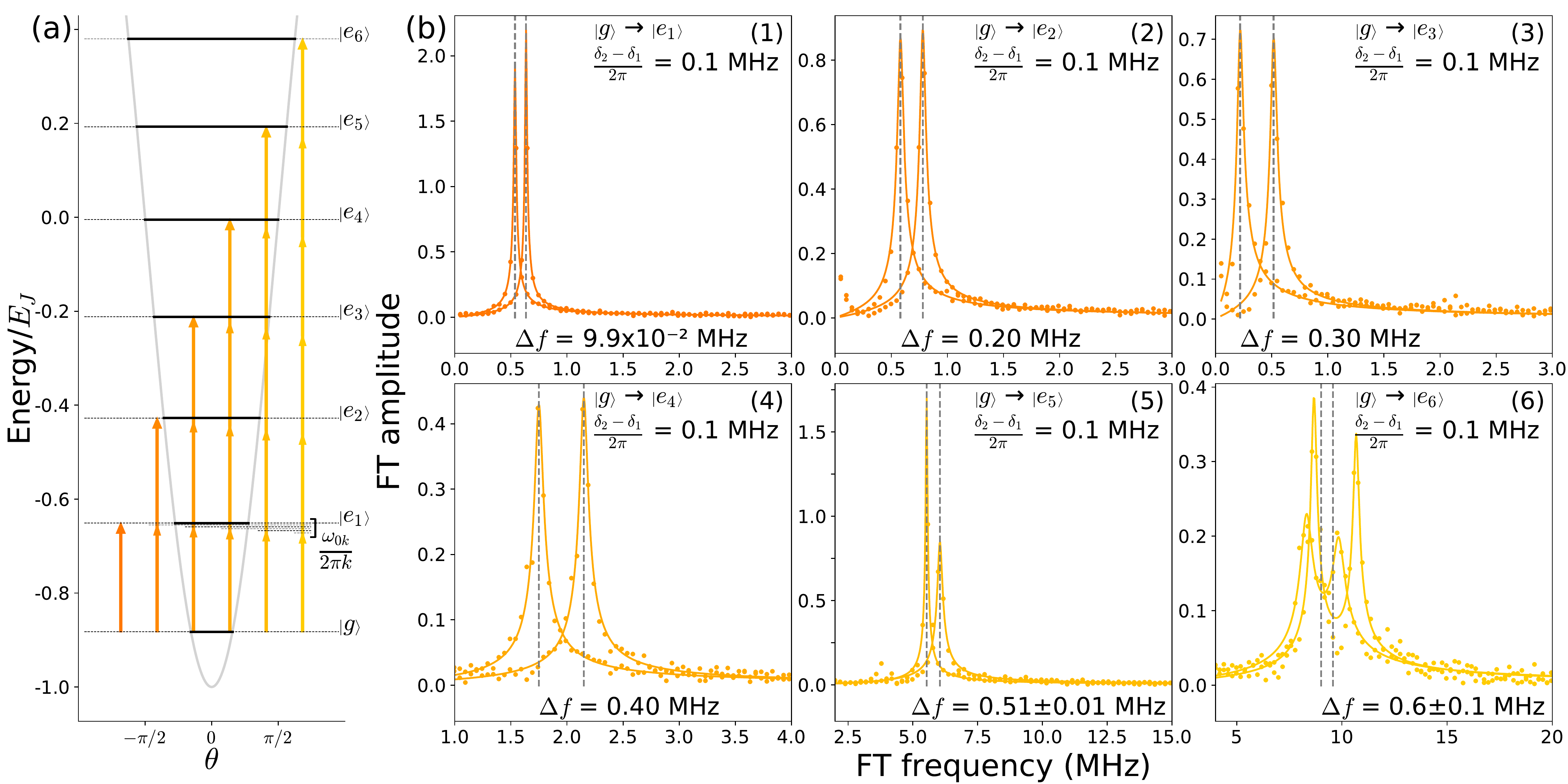}
\caption{Characterization of $k-$photon transitions. (a) Bottom of the cosine potential well of the transmon (grey line) in units of $E_J$ (y-axis), as a function of the phase across the junction (x-axis). The ground state $\ket{g}$ and the first six excited states $\ket{e_{k=1\dots 6}}$ are represented (black horizontal lines). Each excited state $\ket{e_k}$ is prepared with a $k$-photon transition (arrows) with a drive at $\omega_d=\omega_{0k}/k$ (see text). (b1)-(b6) Fourier transform (FT) of a Ramsey signal (dots) for two different drive detunings $\delta_1$ and $\delta_2 = \delta_1 + 2\pi\times 0.1$~MHz. The data is fitted (line) by the Fourier transform of an exponentially decaying cosine function (a sum of two cosines is needed for (b6)), which provides the detuning between the $k$-photon drive with effective frequency $k\omega_d$ and the transition frequency $\omega_{0k}$. We verify that the difference between these detunings $\Delta_f$ scales as $k\left(\delta_2-\delta_1\right)/2\pi$, as expected for a $k-$photon transition. Unlike (b1)-(b5), in (b6) we observe two peaks for each curve. This shows that the Ramsey signal has two frequency components, which is a consequence of quasi-particles tunneling across the junction (see text and Fig.\ref{figs3}).}
\label{figs2}
\end{figure*}

In the main paper, we show that at large pump powers, when the resonator frequency jumps, the transmon is in a highly excited state. We observe the decay of the transmon to its ground state (Fig.~\ref{fig3}b), and verify that the measured resonances coincide with the resonator frequency when the transmon is prepared in excited state $\ket{e_k}$ (Fig.~\ref{fig3}a). In order to drive the transmon from its ground state into state $\ket{e_k}$, we apply a strong pulse at frequency $\omega_{0k}\slash k$ where
\begin{equation}
\omega_{0k} = \frac{E_k-E_0}{\hbar}\;.
\end{equation} 
$E_k$ is the eigenenergy of eigenstate $\ket{e_k}$, and $E_0$ is the eigenenergy of the ground state $\ket{g}$. This induces a $k$-photon transition between $\ket{g}$ and $\ket{e_k}$, as schematically represented in Fig.~\ref{figs2}b. This $k-$photon transition is then used to perform all standard qubit experiments, such as Rabi oscillations between $\ket{g}$ and $\ket{e_k}$, $T_2$ Ramsey and $T_1$ measurements. We verify that this is indeed a $k$-photon transition by performing a Ramsey measurement: two $\pi/2$ rotations with a drive at frequency $\omega_d$ slightly detuned from $\omega_{0k}/k$, separated by a varying time, and followed by the measurement of $\sigma_z$. We obtain the oscillation frequency of the Ramsey fringes by taking the Fourier transform of the measurement (Fig.~\ref{figs2}). If we detune the drive frequency by $\delta$: $\omega_d = \omega_{0k}\slash k - \delta$, then $k\omega_d$ is detuned from the resonance $\omega_{0k}$ by $k\delta$. We perform a first Ramsey experiment with a drive detuned by $\delta_1$, which serves as a reference. A second Ramsey experiment is performed with a detuning $\delta_2 = \delta_1 + 2\pi\times 0.1$~MHz. By subtracting the frequencies of the Ramsey fringes, denoted $\Delta_f$, we get $\Delta_f = k\left(\delta_2 - \delta_1\right)/2\pi$, as shown in Fig.~\ref{figs2}b1-b6.

\subsection{Charge noise}
\label{appendix:chargenoise}
Observing Fig.~\ref{figs2}b, we see that the Ramsey signal is not composed of a single frequency. This is reminiscent of charge noise, where a single quasiparticle is tunneling across the junction in and out of the superconducting island, thus changing its charge parity. This parity change corresponds to changes between $N_g$ and $N_g\pm1/2$, which occur on the millisecond timescale \cite{Riste2013}. Our Ramsey measurements take 16 seconds, hence our data shows the average of these charge offset configurations. Note that despite the fact that we are deep in the transmon regime: $E_J\slash E_C = 140$, the higher energy levels disperse with charge offset. 

\begin{figure*}[h!]
\includegraphics[width=1.5\columnwidth]{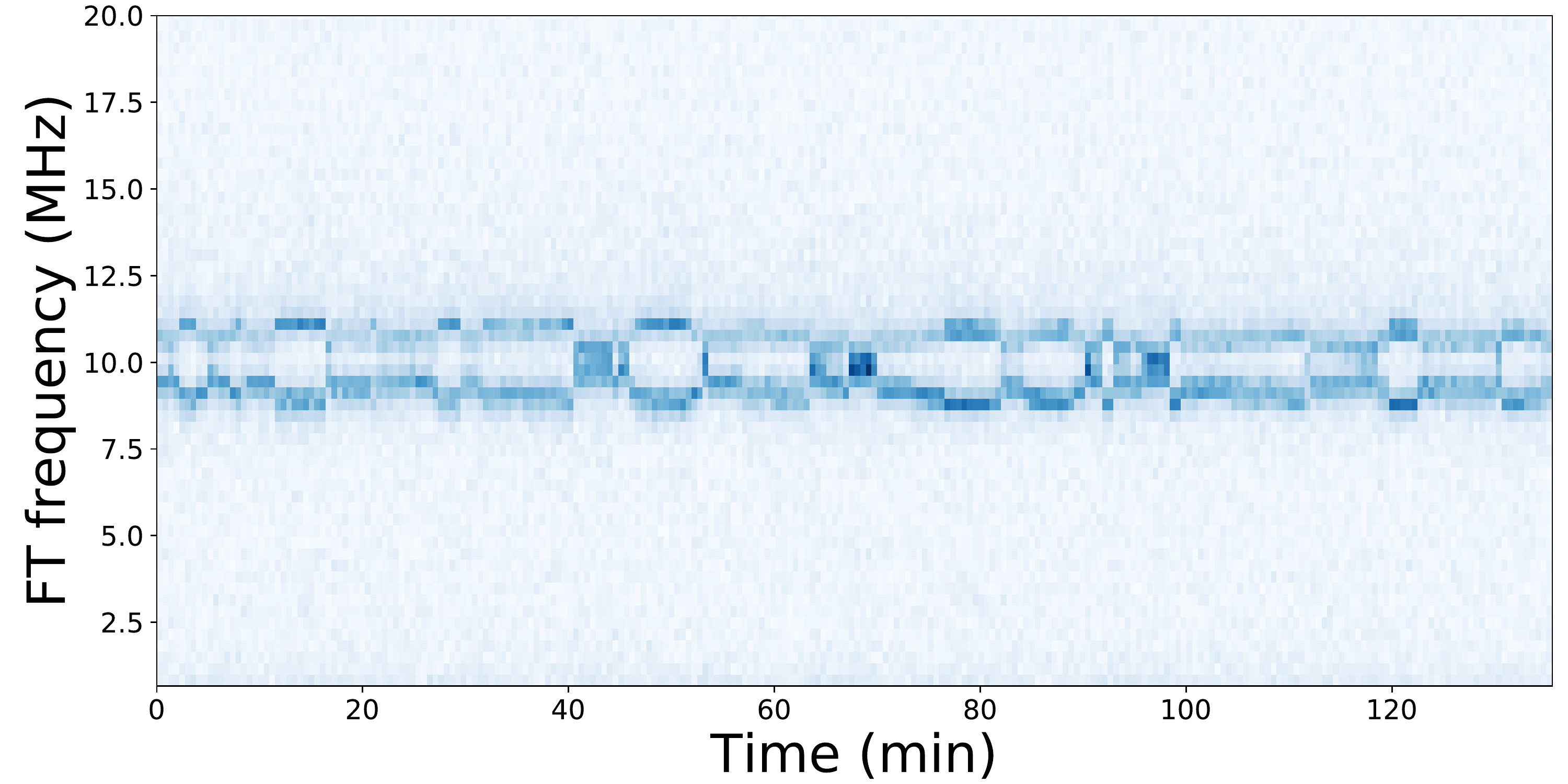}
\caption{Fourier transform (FT) of the Ramsey signal of the $\ket{g}$ to $\ket{e_6}$ transition, measured repeatedly every 32 seconds over a period of about 2 hours. Due to quasiparticle tunneling, two frequencies appear in each Ramsey signal. These frequencies drift over a timescale of several minutes due to background charge motion \cite{Riste2013}.}
\label{figs3}
\end{figure*}

We confirm this effect by plotting the Fourier transform of the Ramsey signal for the $\ket{g}$ to $\ket{e_6}$ transition vs. time (Fig.~\ref{figs3}). We see the slow symmetric drift of the two frequencies, due to the background charge motion \cite{Riste2013}.

The following energy level ($\ket{e_7}$ and above) fluctuate much more than $\ket{e_6}$ with charge offsets, and this is why we could not prepare the transmon in states above $\ket{e_6}$.

\subsection{Numerical simulation}
\label{appendix:numsim}
In this section, we describe the numerical simulation of Fig.~\ref{fig1}c. The details of this simulation can be found in \cite{Verney-al-inpreparation}, and for completeness, we explain the main principles here. We denote $\rho(t)$ the density operator of the transmon-resonator system. This system's dynamics are driven by the time dependent Hamiltonian $\bm{H}(t)$ of Eq.~\eqref{hamiltonian}, and by a coupling to a bath through an operator $\bm{X}$. Typically, $\bm{X} = \bm{a}+\bm{a}^\dag$, and the bath is a transmission line. For simplicity, we neglect the direct coupling of the transmon to the bath, but this can easily be included in the theory by changing $\bm{X}$. We recall that $\bm{H}(t)$ is periodic with period $T_p = 2\pi/\omega_p$, where $\omega_p$ is the pump frequency. The goal is to find $\rho(t)$ in the infinite time limit. 

After a sufficiently long time, $\rho(t)$ reaches a limit-cycle: a $T_p$-periodic trajectory denoted $\bar\rho(t)$. We find this limit-cycle using Floquet-Markov theory \cite{Grifoni-Hanggi-PhysRep_1998}. This theory assumes a weak coupling to the bath, but can treat pumps of arbitrary amplitudes and frequencies.

\paragraph{Change of frame:} The first step is to write the Hamiltonian in a frame where the resonator is close to the vacuum state. We do this by solving the quantum Langevin equations when $E_J=0$. We will use the following useful formulas \cite[(4.12)]{Devoret1995}:
\begin{equation*}
[\bm{\theta},\bm{N}] = i\;,
\end{equation*}
and hence for any analytic function $f$
\begin{equation*}
[\bm{\theta},f(\bm{N})] = i\frac{\partial{f}}{\partial \bm{N}}(\bm{N})\;,
\end{equation*}
and also
\begin{equation*}
[\bm{a},f(\bm{a}^\dag)] = \frac{\partial{f}}{\partial \bm{a}^\dag}(\bm{a}^\dag)\;.
\end{equation*}
We denote $\bm{a}_0, \bm{N}_0, \bm{\theta}_0$ the solutions of the Langevin equations associated to Hamiltonian \eqref{hamiltonian} with $E_J=0$. We find:
\begin{eqnarray*}
\frac{d}{dt}\bm{a}_0 &=&\frac{1}{i\hbar}[\bm{a}_0,\bm{H}] = -i\omega_0\bm{a}_0-ig\bm{N}_0-i\mathcal{A}_p(t)\\
\frac{d}{dt} \bm{N}_0&=&\frac{1}{i\hbar}[\bm{N}_0,\bm{H}] = 0\\
\frac{d}{dt} \bm{\theta}_0&=&\frac{1}{i\hbar}[\bm{\theta}_0,\bm{H}] =8E_C/\hbar\bm{N}_0+g(\bm{a}_0+\bm{a}_0^\dag)\;,
\end{eqnarray*}
where $\mathcal{A}_p(t)=A_p\cos(\omega_pt)$, and $A_p\ge 0$.
These equations are linear, and are therefore easy to solve. We introduce the expectation values: $\braket{\bm{a}_0} = a_0, \braket{\bm{N}_0} = N_0, \braket{\bm{\theta}_0} = \theta_0$. Using some approximations \cite{Verney-al-inpreparation}, we find
\begin{eqnarray*}
{a}_0(t) &=& \frac{A_p/2}{(\omega_p-\omega_a)}e^{-i\omega_p t}\\
N_0(t) &=& 0\\ 
\theta_0(t)&=&g\frac{A_p}{\omega_p(\omega_p-\omega_a)}\sin(\omega_pt)\;.
\end{eqnarray*}
We define
\begin{equation}
\label{nbar}
\bar{n} = \left|\frac{A_p/2}{\omega_p-\omega_a}\right|^2\;.
\end{equation}
For simplicity we assume $\omega_p>\omega_a$. If $\omega_p<\omega_a$ then one needs to multiply the following solutions by $-1$. We get
\begin{eqnarray}
\label{sol0}
{a}_0(t) &=& \sqrt{\bar{n}}e^{-i\omega_p t}\notag\\
N_0(t) &=& 0\\ 
\theta_0(t)&=& 2\frac{g}{\omega_p}\sqrt{\bar{n}}\sin(\omega_pt)\;.\notag
\end{eqnarray}
Comparing Eq.~\eqref{nbar}, and Eq.~\eqref{nbar_r}, we see that the expressions of $\bar n$ and $\bar n_r$ are very similar, and differ only by the presence of subscripts $r$, and the term in $\kappa_r$ in Eq.~\eqref{nbar_r}. In the limit of large detunings: $|\omega_r - \omega_p|\gg \kappa_r$, and hence the term in $\kappa_r$ can be neglected. The pump amplitude $A_{p,r}$ slightly differs from $A_p$ due to the transmon-resonator mode hybridization. For our experimental parameters, $A_{p,r}$ and $A_p$ differ by less than $1\%$. The measured resonator frequency $\omega_r$ roughly varies between $\bar\omega_r$ and $\omega_a$, spanning about $20$~MHz. For the simulation of Fig. \ref{fig1}b, we consider a pump at frequency $\omega_p/2\pi=8.1$~GHz, which is detuned from the resonator by about $340$~MHz. This detuning is much larger than the frequency variation of $\omega_r$. Hence $\bar n$ and $\bar n_r$ differ by less than $15\%$.

We now move to a frame around the solutions given in Eq.~\eqref{sol0}, introducing $\tilde{\bm{\theta}}=\bm\theta-\theta_0$, $\tilde{\bm{N}}=\bm{N}-{N}_0$ and $\tilde{\bm{a}}=\bm{a}-{a}_0$. This frame is chosen such that the deviation of the resonator from the vacuum is small, and hence we can use a low truncation for the resonator mode. This change of frame leads to the following Hamiltonian \cite{Verney-al-inpreparation}:

\begin{eqnarray*}
\tilde{\bm{H}}&=&4E_C \tilde{\bm{N}}^2-E_J\cos\left(\tilde{\bm{\theta}}+\theta_0(t)\right)
\\
&+&\hbar\omega_a\tilde{\bm{a}}^\dag \tilde{\bm{a}}\notag + \hbar g\tilde{\bm{N}}\left(\tilde{\bm{a}}+\tilde{\bm{a}}^\dag\right)\\
&=&4E_C \tilde{\bm{N}}^2-E_J\left(\cos(\tilde{\bm{\theta}})\cos(\theta_0(t))-
\sin(\tilde{\bm{\theta}})\sin(\theta_0(t))\right)\\
&+&\hbar\omega_a\tilde{\bm{a}}^\dag \tilde{\bm{a}}\notag + \hbar g\tilde {\bm{N}}\left(\tilde{\bm{a}}+\tilde{\bm{a}}^\dag\right)
\end{eqnarray*}

We write $\tilde{\bm{H}}$ as a matrix in the basis spanned by $\ket{T_k}\otimes \ket{F_l}$, where $\ket{T_k}$ are eigenstates of the transmon Hamitlonian $4E_C \tilde{\bm{N}}^2-E_J\cos\left(\tilde{\bm{\theta}}\right)$, and $\ket{F_l}$ are Fock states: eigenstates of the resonator Hamiltonian $\hbar\omega_a\tilde{\bm{a}}^\dag \tilde{\bm{a}}$. We truncate this basis to $M=450$ states, with $45$ states for the transmon and $10$ states for the resonator.

\paragraph{Floquet analysis:} Using functions provided by the Quantum Toolbox in Python (Qutip), we compute the Floquet states of $\tilde {\bm{H}}(t)$, which we denote $\ket{\psi_\alpha(t)}$ for $\alpha = 1,\dots, M$. We can then calculate the dissipation operator $\bm{X}$ matrix elements in the Floquet basis $\ket{\psi_\alpha}$ \cite[(245)]{Grifoni-Hanggi-PhysRep_1998}.

With these matrix elements, we can now calculate the master equation followed by the density matrix \cite[(251)]{Grifoni-Hanggi-PhysRep_1998}, and compute its steady state in the Floquet basis: $\bar\rho(t) = \sum_\alpha{\rho_{\alpha\alpha}\ket{\psi_\alpha(t)}\bra{\psi_\alpha(t)}}$. We recall that $\bar\rho(t)$ is $T_p$-periodic and we define $\bar\rho_0=\bar\rho(0)$. We can express $\bar\rho_0$ in the basis $\{\ket{T_k}\otimes \ket{F_l}\}_{k,l}$, and calculate the partial trace with respect to the cavity $\bar{\rho}_{0q} = \text{Tr}_\text{cav}(\bar{\rho}_0)=\sum_l{\bra{F_l}\bar\rho_0\ket{F_l}}$. The diagonal of $\bar\rho_{0q}$: diag$(\bar\rho_{0q})$ is the list of the transmon eigenstate populations. Note that at times $t_m = 0 + m T_p$, $\theta_0(t_m)=0$, hence the unitary which transforms states between the displaced frame back to the original one is the identity for the transmon degree of freedom. In Fig. \ref{fig1}c, we plot diag$(\bar\rho_{0q})$ as a function of $\bar{n}$, where $\bar{n}$ given in Eq.~\eqref{nbar}.


\subsection{Quasiparticle generation}
\label{appendix:quasiparticles}
\begin{figure*}[h!]
\includegraphics[width=1.5\columnwidth]{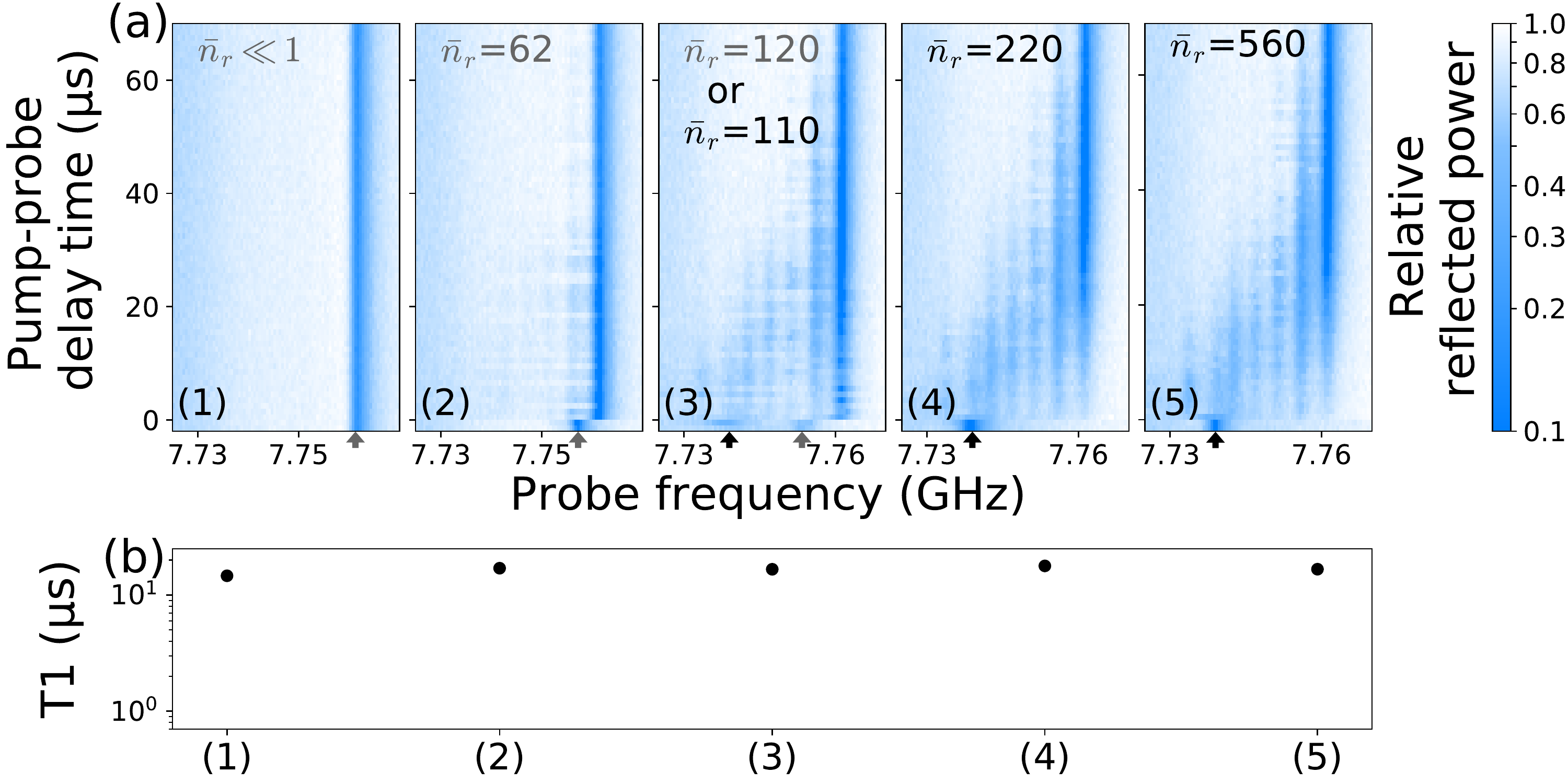}
\caption{Transmon relaxation following a pump pulse of increasing power. (\textbf{a}) Same experiment as in Fig.~\ref{fig3}b for increasing pump power from a1 to a5, a5 corresponding to the maximum power the microwave source could deliver. (\textbf{a1}-\textbf{a2}) The pump induces an AC Stark shift proportional to $\bar{n}_r$ at $t<0$ but does not significantly excite the transmon. (\textbf{a3}) At intermediate pump powers, the system is in a statistical mixture of two configurations, as indicated by the two dips at $t<0$. One corresponds to the AC Stark shifted cavity, the other to the bare cavity frequency $\omega$ indicating that the transmon gets highly excited. (\textbf{a4}-\textbf{a5}) At higher pump powers, the cavity frequency always jumps to $\omega$ and the transmon decays as in Fig.~\ref{fig3}b. (\textbf{b}) Transmon $T_1$ (y-axis) measured 50$~\mu$s after a pump is applied. The pump duration and powers (x-axis) correspond to the ones used for the experiments shown in (a). $T_1$ does not decrease as the pump power increases which indicates that the pump does not generate a measurable increase in the quasiparticle density \cite{Wang-al-NatureComm_2014, Vool-al-PRL_2014}.}
\label{figs4}
\end{figure*}

In section \ref{sec:quasiparticles}, we discuss whether the strong pump we apply on the system generates quasiparticles. If the pump significantly increases the quasiparticle density, the transmon $T_1$ will decrease \cite{Wang-al-NatureComm_2014, Vool-al-PRL_2014}. In Fig.~\ref{figs4}a, we reproduce the experiment of Fig.~\ref{fig3}b for various pump powers, which correspond to various numbers of photons $\bar{n}_r$. For small $\bar{n}_r$, the transmon is not excited, while for large $\bar{n}_r$, the transmon is driven to highly excited states. Additionally, after each pump pulse, we measure the transmon $T_1$ after a time delay chosen so that all modes have decayed back to their ground state. We find no decrease in $T_1$ (Fig.~\ref{figs4}b) which means that the pump can expel the transmon out of its potential well without producing a measurable amount of quasiparticles.

\bibliography{biblio}

\end{document}